\documentclass[a4paper]{article}
\usepackage{amsmath}
\usepackage[english]{babel}
\usepackage{latexsym}
\usepackage{amssymb}
\usepackage{amscd}
\usepackage{amsgen,amstext,amsbsy,amsopn}
\usepackage{math rsfs}

\newcommand{\version}{August 30, 2008}
\numberwithin{equation}{section}
\newcommand{\bdm}{\begin{displaymath}}
\newcommand{\edm}{\end{displaymath}}
\newcommand{\bdn}{\begin{eqnarray}}
\newcommand{\edn}{\end{eqnarray}}
\newcommand{\bay}{\begin{array}{c}}
\newcommand{\eay}{\end{array}}
\newcommand{\ben}{\begin{enumerate}}
\newcommand{\een}{\end{enumerate}}
\newcommand{\beq}{\begin{equation}}
\newcommand{\eeq}{\end{equation}}
\newcommand{\bml}[1]{\begin{multline} #1 \end{multline}}

\newcommand{\lf}{\left}
\newcommand{\ri}{\right}

\newcommand{\N}{\mathbb{N}}

\newcommand{\Z}{\mathbb{Z}}

\newcommand{\xv}{\vec{x}}
\newcommand{\xvp}{\vec{x}^{\prime}}
\newcommand{\rv}{\vec{r}}

\newcommand{\diff}{\mathrm{d}}
\newcommand{\eps}{\varepsilon}

\newcommand{\bo}{\mathcal{B}_1}

\newcommand{\gpf}{\mathcal{E}^{\mathrm{GP}}}
\newcommand{\gpd}{\mathcal{D}^{\mathrm{GP}}}
\newcommand{\gpe}{E^{\mathrm{GP}}}
\newcommand{\gpm}{\Psi^{\mathrm{GP}}}
\newcommand{\tgpf}{\tilde{\mathcal{E}}^{\mathrm{GP}}}
\newcommand{\tgpe}{\tilde{\mathcal{E}}^{\mathrm{GP}}}
\newcommand{\gpfi}{\mathcal{E}^{(i)}}
\newcommand{\gpfit}{\tilde{\mathcal{E}}^{(i)}}

\newcommand{\glf}{\mathcal{E}^{\mathrm{GL}}}
\newcommand{\tff}{\mathcal{E}^{\mathrm{TF}}}

\newcommand{\tfe}{E^{\mathrm{TF}}}

\newcommand{\tfm}{\rho^{\mathrm{TF}}}
\newcommand{\rin}{R_{\mathrm{h}}}
\newcommand{\supp}{\mathrm{supp}\lf(\tfm\ri)}
\newcommand{\supph}{\mathcal{T}_{}}
\newcommand{\const}{C}

\newcommand{\trial}{{\Psi}}

\newcommand{\phase}{g}

\newcommand{\latt}{\mathcal{L}}

\newcommand{\spac}{\ell}

\newcommand{\celli}{\mathcal{Q}^i_{}}

\newcommand{\ucell}{\mathcal{Q}_1}
\newcommand{\numb}{N}

\newcommand{\chem}{\mu^{\mathrm{GP}}}
\newcommand{\tfchem}{\mu^{\mathrm{TF}}_{}}

\newcommand{\tfma}{\rho^{\mathrm{TF}}}

\newcommand{\magnp}{\vec{A}}
\newcommand{\rmagnp}{\vec{B}}

\newcommand{\half}{\hbox{$\frac12$}}

\newcommand{\rtf}{R_{\rm h}}
\newcommand{\bii}{\mathcal{B}_{\varepsilon}^i}

\newtheorem{teo}{Theorem}[section]
\newtheorem{lem}{Lemma}[section]
\newtheorem{pro}{Proposition}[section]

\newenvironment{proof1}[1]{\mbox{} \newline \noindent \textbf{Proof of #1:} \newline \noindent}{\begin{flushright} $ \Box $ \end{flushright}}
\newenvironment{proof}{\emph{Proof:}}{\begin{flushright} $ \Box $ \end{flushright}}

\begin{document}

\markboth{\scriptsize{CY \version}}{\scriptsize{CY \version}}

\title{Energy and Vorticity\\ in Fast Rotating Bose-Einstein Condensates}

\author{M. Correggi	\\ \normalsize\it Scuola Normale Superiore SNS,    \\ \normalsize\it Piazza dei Cavalieri 7, 56126 Pisa, Italy.	\\ \normalsize\it \hspace{-.5 cm}	\\ J. Yngvason	\\ \normalsize\it Erwin Schr{\"o}dinger Institute for Mathematical Physics,	\\ \normalsize\it Boltzmanngasse 9, 1090 Vienna, Austria,	\\ \normalsize\it Fakult\"at f\"ur Physik, Universit{\"a}t Wien,	\\ \normalsize\it Boltzmanngasse 5, 1090 Vienna, Austria.}

\date{\version}

\maketitle

\begin{abstract}
	We study a  rapidly rotating Bose-Einstein condensate confined to a finite  trap in the framework of two-dimensional Gross-Pitaevskii theory in the strong coupling (Thomas-Fermi) limit. Denoting the coupling parameter by $1/\eps^2$ and the rotational velocity by $\Omega$,  we evaluate exactly  the next to leading order contribution to the ground state energy in the parameter regime 
	$|\log\eps|\ll \Omega\ll 1/(\eps^2|\log\eps|)$ with $\eps\to 0$. While the TF energy includes only the contribution of the centrifugal forces the next order corresponds to a lattice of vortices whose density is proportional to the rotational velocity. 
	\vspace{0,2cm}

	\vspace{0,2cm}
	
\end{abstract}

\section{Introduction}

Bose-Einstein condensates respond to rotational motion of the enclosing container by the creation of quantized vortices. This remarkable manifestation of superfluidity has been studied, both experimentally and theoretically, in dilute, ultracold Bose gases since almost a decade and still offers a number of unsolved problems. We refer to the monograph \cite{Afta}, the review article \cite{FetterRMP}, as well as the papers \cite{Fetter}-\cite{CDY2} for extensive lists of references.
Most of the theoretical work has been carried out within the framework of the Gross-Pitaevskii (GP) equation for the wave function of the condensate. In the GP equation the interaction is encoded in a single parameter $g=4\pi Na/L$,  where $a$ is the scattering length of the interaction potential, $N$  the particle number and $L$ the length scale associated with the external confining potential. For rotating gases in their ground state  the GP equation was derived in \cite{Seir2} (upper bound) and  \cite{Lieb1} (lower bound)
from the quantum mechanical many-body Hamiltonian with  purely repulsive, short range interactions and fixed values of the rotational velocity and the coupling parameter as $N\to\infty$.  The extension of this derivation to the case when the coupling parameter and the rotational velocity tend to infinity (or approach a critical value in the case of harmonic traps) has not yet been completed,  but the leading order asymptotics  of the many-body energy for large coupling and rotational velocity in anharmonic traps was computed in \cite{BCPY}.

Detailed results on the emergence of vortices as the rotational velocity is increased have been obtained within two-dimensional GP theory when the GP interaction parameter is large (\lq Thomas-Fermi' limit) and the rotational velocity is of the order of the logarithm of this parameter \cite{ AftaDu, Seir1, Igna1, Igna2, RD}.  In this case the number of vortices remains finite as the interaction parameter tends to infinity. The rotation has no effect on the energy to leading order in the coupling parameter but there is  a logarithmic contribution due to the vortices in the next to leading order. By contrast, the papers \cite{FiB}--\cite{CDY2} are  mainly concerned with the situation when the rotation is so fast that the centrifugal energy and the interaction energy are comparable in magnitude.  This holds when the rotational velocity increases like the square root of the interaction parameter. This case was also considered in the many-body context in \cite{BCPY}, relying partly on estimates from \cite{CDY1}.  
The main result of \cite{CDY1} and  \cite{CDY2} was the rigorous  evaluation of the GP ground state energy  to leading order in the interaction parameter in the regime just mentioned. This energy can be computed by minimizing a simple density functional that  contains besides the interaction term another term of the same order corresponding to the centrifugal potential. This functional was first introduced 
 in \cite{FiB} where many of the basic insights about the physics of rapidly rotating condensates in anharmonic traps can be found, see also \cite{KB} and \cite{B}. We note that,
since the rotational velocity is unbounded,  the confining potential must increase more rapidly than quadratically with the distance form the rotational axis in order that the centrifugal forces do not tear the condensate apart. 

In the present paper we evaluate exactly the next term in the asymptotic expansion beyond the leading contribution in  the parameter regime 
\beq
\label{cond omega}
	|\log\eps| \ll \Omega \ll {1}/({\eps^{2} |\log\eps|})
\eeq
where the coupling parameter $g$ has been written as $1/\eps^2$ with $\eps\to 0$, and $\Omega$ is the rotational velocity. We remark that the dimensionless parameter $\varepsilon$ can be interpreted as the ratio between the \lq healing length' $(4\pi N/L^3)^{-1/2}$ and the extension $L$ of the confining trap.
The subleading term  in the energy corresponds to the energy of a lattice of vortices of degree  one such that the total vorticity is proportional to the rotational velocity.  In order to bring out the salient points as simply as possible we restrict ourselves to the model considered in \cite{CDY1}, i.e., the case of a flat, circular trap (\lq bucket') of finite radius. 

When computing the upper bound on the energy we make a variational ansatz with a wave function that is essentially the product of  a shape function, taking the deformation due to the centrifugal forces into account,  and a function corresponding to a lattice of vortices uniformly distributed over the trap. The evaluation of the energy can be cast in the form of an electrostatic problem with the vortices playing the role of point charges while the vector potential due to the rotation can be regarded as an electric field generated by a uniform charge distribution. The optimal arrangement of the vortices is then determined by  a minimization problem for the total electrostatic energy. When the rotational velocity reaches  $O(1/(\eps^2|\log\eps|))$ a different trial function, with the vorticity concentrated in a region where the density is small (\lq giant vortex'),  gives a lower energy. This transition was first noted  in \cite{FiB} and the estimates of the present paper corroborate it since our rigorous upper bound to the energy is smaller than the energy of the giant vortex  if $\Omega\ll 1/(\eps^2|\log\eps|))$.

To prove the lower bound the problem is reformulated in such a way that results from  Ginzburg-Landau (GL) theory obtained in \cite{SS1} and \cite{SS2} can be employed. The strong inhomogeneity of the density in fast rotating condensates causes  problems that make the reduction to the GL case not entirely straightforward, but once these have been overcome  a lower bound that matches the upper bound to subleading order in the asymptotic parameter range \eqref{cond omega} can be derived.
The techniques of \cite{SS1} and \cite{SS2} also turn out to be useful for the investigation of the vorticity of the minimizer.

\section{The Mathematical Setting}

We now recall the setting of \cite{CDY1}  that will be used in the present paper. The condensate is confined to the two-dimensional  unit disc $\bo$ and the rotational axis  is perpendicular to the disc and passes through its center. We note that  this model  can also be applied to the description of a three-dimensional rotating condensate confined to a long cylinder. The plane of the disc is the $xy$ plane and $\vec r=(x,y)$ is the position vector with length $r$, while  $\vec{e}_z$ denotes the unit vector in the $z$ direction. The complex valued  order parameter (the wave function of the condensate)  is denoted by  $\Psi(\vec r)$.  In the non-inertial rotating frame the GP energy functional 
can be written as
\beq
	\label{GPfunctional}
	\gpf[\Psi] = \int_{\bo} \diff \rv \: \lf\{ \lf|\lf(  \nabla - i \magnp \ri) \Psi \ri|^2 - \frac{\Omega^2 r^2 |\Psi|^2}{4} + \frac{|\Psi|^4}{\eps^2} \ri\},
\eeq
where the vector potential $ \magnp $ is given by 
\beq	
	\magnp \equiv \frac{\Omega}{2} \: \vec{e}_z \wedge \rv.
\eeq
For convenience we also introduce the abbreviation
\beq
	\omega \equiv \eps \Omega. 
\eeq
For fixed $\omega$ the centrifugal and the interaction terms in \eqref{GPfunctional} are both $O(1/\eps^2)$. The kinetic first term, containing $A \sim \Omega$, is formally also of order $1/\eps^2$ if $\Omega\sim 1/\eps$, but a complex phase factor in $\Psi$, due to vortices, can partly compensate the effect of $\vec A$. Indeed, in the ground state this term is of lower order as we shall see.

The ground state properties of the condensate are obtained by minimizing the GP functional over the domain
\beq\label{GPdomain}
	\gpd = \lf\{ \Psi \in H^1(\bo) \: | \: \| \Psi \|_2 = 1 \ri\}.
\eeq
Here $H^1(\bo)$ denotes the Sobolev space of complex valued functions $\Psi$ on $\bo$ such that both $\Psi$ and $\nabla \Psi$ are square integrable.
The choice (\ref{GPdomain}) naturally leads to (magnetic) Neumann boundary conditions for the minimizer on $\partial\mathcal B_1$.  
Alternatively one could impose Dirichlet  boundary conditions.  For $\Omega\sim 1/\eps$ this would affect the energy to order $O(1/\eps)$ that is negligible compared to the vortex contribution $O(\Omega |\log \eps|)$ that we are interested in. In other parameter regions the effect of the boundary conditions can be more significant, and the same remark applies to an extension of our analysis to homogeneous potentials as in \cite{CDY2}. For simplicity we shall, however, in this paper stick to the choice \eqref{GPdomain} that highlights the vortex contributions.

We denote by $ \gpe $ the GP ground state energy and by $ \gpm $ any corresponding minimizer. The existence of such minimizer(s)  as well as the fact that any minimizer solves  the GP differential equation
\beq
	\label{GP Equation}
	- \lf( \nabla - i \magnp \ri)^2 \gpm - A^2 \gpm + 2 \eps^{-2} \lf| \gpm \ri|^2 \gpm = \chem \gpm,
\eeq
with boundary condition $ \nabla_r \gpm = 0 $ on $ \partial \bo $ can be deduced by standard techniques (see, e.g., \cite{Seir1}). The chemical potential $ \chem $ is fixed by the $L^2-$normalization of $ \gpm $, i.e., \beq\chem = \gpe + \eps^{-2} \| \gpm \|_4^4 .\eeq

In \cite{CDY1} we studied the asymptotics of $ \gpe $ as $ \eps \to 0 $  and proved that the energy is well approximated to  leading order by minimizing  the \lq Thomas-Fermi' (TF)  functional
\beq
	\label{TFfunctional}
    	\tff[\rho] = \frac1{\eps^2}\int_{\bo} \diff\rv \: \left\{ \rho^2 - \frac{\omega^2 r^2 \rho}{4} \right\}.
\eeq
Note that, unlike in \cite{CDY1, CDY2}, we have  included the factor $1/\eps^2$ in the definition of $\tff$. The density $ \rho(\rv)\geq 0 $ is the probability density associated with a condensate wave function $ \Psi $, i.e., $ \rho = |\Psi|^2 $. The TF ground state energy,
\beq
	\label{tfe}
    	\tfe \equiv \min_{\| \rho \|_1 = 1, \rho\geq 0} \tff[\rho] = \tff[\tfm],
\eeq
and the corresponding normalized density $ \tfm $ can be explicitly calculated.  The formulas and some properties of relevance for this paper are collected in the Appendix. We note in particular that the centrifugal forces may create a \lq hole' in $ \tfm $, i.e., the density $ \tfm $ vanishes on a  disc centered at the origin if $\omega>\omega_{\rm h}\equiv 4/\sqrt{\pi}$. 

\section{The Main Results}
The main results proved in \cite{CDY1} are, in a slightly different notation,  contained  in the following 

\begin{teo}[Leading order ground state energy asymptotics \cite{CDY1}]
	\label{energyold}
	\mbox{}	\\
	If $ \Omega\sim 1/\eps $ as $ \eps \to 0 $, then
	\beq
		\label{energyasy}
		 \gpe = \tfe + {O}\lf( \eps^{-1} |\log \eps| \ri),
	\eeq
	whereas
		 if $ 1/\eps\ll\Omega $,
	\beq
		\label{energyasyu}
		 \gpe = \tfe + O(\eps^{-2})+O((\eps\Omega)^2|\log\eps|).
	\eeq
\end{teo}

In this paper we investigate the correction beyond the leading order TF  term for the parameter range
$|\log\eps|\ll\Omega\ll1/(\eps^2|\log\eps|)$. Our main result is as follows (the notation $\Omega\lesssim 1/\eps$ means that  $\Omega\leq C/\eps$ as $\eps\to 0$, with some $C<\infty$):

\begin{teo}[Improved ground state energy asymptotics]
	\label{energy}
	\mbox{}	\\
	If  $ |\log\eps| \ll \Omega \lesssim\eps^{-1} $ as $ \eps \to 0 $, then 
	\beq
		\label{energyub1}
        	\gpe = \tfe +\frac{ \Omega |\log(\eps^2 \Omega)|}{2} (1+o(1)),
	\eeq
	whereas, if $ \eps^{-1} \lesssim \Omega \ll \eps^{-2} |\log\eps|^{-1} $,
	\beq
		\label{energyub2}
		 \gpe = \tfe + \frac{ \Omega |\log\eps|}{2}  (1+o(1)).
	\eeq
\end{teo}

In \cite{CDY1} we have shown that as a consequence of the energy asymptotics $|\gpm|^2 $ converges as $ \eps \to 0 $ to $ \tfm $ in $L^1$-norm. Inside the hole, if present, it is exponentially small, i.e., bounded by $\exp(-{\rm const.}/\eps^\beta)$ for a $\beta>0$. See \cite{CDY1}, Propositions 2.4 and 2.5.

The energy bounds also allow to prove a result about the uniform distribution of the vorticity of
$\Psi^{\rm GP}$ outside the hole, at least for $\Omega\lesssim 1/\eps$:

\begin{teo}[Uniform distribution of vorticity]
	\label{distribution}
	\mbox{}	\\
	Let $ \gpm$ be any GP minimizer and $ \eps > 0 $ sufficiently small. If $ |\log\eps| \ll \Omega \lesssim \eps^{-1} $, there exists a finite family of disjoint balls $ \lf\{ \bii \ri\} \subset \supp $ such that 
	\ben
		\item	the radius of any ball is smaller than $ \Omega^{-1/2} $, 
		\item	the sum of all the radii is much smaller than $ {\Omega}^{1/2} $, 
		\item	on $ \partial \bii $,
$ \lf| \gpm \ri| \geq \const |\log(\eps^2\Omega)|^{-1} $ with $C>0$
\een
	and, denoting by $ \rv_{i,\eps} $ the center of each ball $ \bii $ and by $ d_{i,\eps} $ the winding number
		of $ |\gpm|^{-1} \gpm $ on $ \partial \bii $,
	\beq
	\label{measure}
		\frac{2 \pi}{\Omega} \sum d_{i,\eps} \delta\lf(\rv - \rv_{i,\eps} \ri) \: \underset{\eps \to 0}{\overset{\mathrm{w}}{\longrightarrow}} \:\: \chi^{\mathrm{TF}}(\rv) \: \diff \rv,
	\eeq
	in the sense of measures, where $ \chi^{\mathrm{TF}}(\rv) $ stands for the characteristic function of $ \supp $. 
\end{teo}

For $\Omega\gg 1/\eps$ the vorticity distribution is still an open question. In this regime $ \gpm $ is not uniformly bounded in $ \eps $ and is essentially supported in an annulus of very small width $\sim \omega^{-1}$ close to the boundary. As we shall see, a trial function with a uniform distribution of vortices still gives the right energy to subleading order for $\Omega\ll 1/(\eps^2|\log\eps|)$, but for larger $\Omega$ a trial function without any vortices in the support of $\rho^{\rm TF}$ (a \lq giant vortex') has lower energy. It can be expected that for the true minimizers the vortices are gradually expelled from the essential support of the density as $\Omega$ approaches $1/(\eps^2|\log\eps|)$ but there are so far no rigorous results on the details of this phenomenon. The numerical investigations of \cite{KTU}, however, support this picture.

\section{Energy Upper Bound}

For an upper bound we test the functional  \eqref{GPfunctional} with a trial function of the form
        \beq
        \label{trial function}
           \Psi_{}(\rv) = c_{} \sqrt{\rho_{}(\rv)} \: \xi_{}(\rv) \phase(\rv),
        \eeq
        where $c_{}$ is a normalization constant, $ \rho_{} $ a suitable regularization of $ \rho^{\rm TF} $, $ \phase $ is a phase factor, and  $\xi$ a  function that vanishes at the vortices, i.e., the singularities of $ \phase $. 
To define the functions precisely we first  introduce some notation.
\newline
We denote by $\mathcal L_{}$ a finite, regular lattice (triangular, rectangular or hexagonal) of points $\rv_i\in\mathcal B_1$. Each lattice point  $\rv_i$ lies at  the center of a lattice cell $Q_{}^i$ and the lattice constant $\ell$ is chosen so that the area of  $Q_{}^i$ is 
\beq
|Q_{}^i|= \frac{2\pi}{\Omega}.\eeq
Thus, 
\beq \ell=\hbox{\rm (const.)}\Omega^{-1/2} \eeq
and the total number of lattice points in the unit disc is
\beq {\mathcal N}=\frac{\Omega}{2}(1+O(\Omega^{-1/2})).\eeq
For large $\omega=\eps\Omega$ the support of $\rho^{\rm TF}$ has an area of the order $(\omega+1)^{-1}$ and the number of lattice points on the support of $\rho^{\rm TF}$ is of the order
\beq\label{vortinside}{ \mathcal N}'=(\omega+1)^{-1}\Omega.\eeq
In particular, for $\Omega\gg 1/\eps$, ${ \mathcal N}'=O(1/\eps)$.

Using complex notation $\zeta=x+iy$ for the points $\rv=(x,y)\in\mathbb R^2$ the phase factor $g_{}$ is defined as
\beq
    	\label{phase}
        	\phase(\rv) = \prod_{\zeta_i \in \latt} \frac{\zeta - \zeta_i}{|\zeta- \zeta_i|}.
    	\eeq
The phase factor is singular at the lattice points but these singularities are compensated by the function
\beq
	\xi_{}(\rv)=	
		\lf\{
		\begin{array}{ll}
            		1				&	\mbox{if} \:\: |\zeta - \zeta_i| \geq t,  \\
            		t^{-1} |\zeta - \zeta_i|	&   	\mbox{if} \:\: |\zeta - \zeta_i| \leq t.
		\eay
		\ri.  	
\eeq
Here $t$, with 
\beq
	\label{tcondition}
	\min \{ \eps, ({\eps}/{\Omega})^{1/2} \} \leq t \ll \Omega^{-1/2},
\eeq 
is a variational parameter that will be fixed later. Thus $\xi_{}(\rv)$ vanishes at the lattice points $\rv_i$ and is equal to 1 outside of the union of the discs $\mathcal B_{t}^i$ of radius $t$ centered at those points. 

The size of $t$ can be estimated by the following heuristic argument. The kinetic energy of a vortex in a cell is of the order
$e_{\rm kin}\sim \int_t^\ell (1/r)^2 r\,dr\sim \log(\ell/t)$.
Creating a vortex also causes an excess interaction energy because the density depletion in the vortex core of radius $t$ has to be compensated by an increase in density elsewhere. This leads to the additional interaction energy
$e_{\rm int}\sim \rho\,(t/\eps)^2$, 
and by minimizing $e_{\rm kin}+e_{\rm int}$ we obtain $t\sim \eps\rho^{-1/2}$. For slow rotations where $\rho=O(1)$ this gives $t\sim \eps$, while for rapid rotation, where $\rho=O(\omega)$, we obtain $t\sim (\eps/\Omega)^{1/2}$. These heuristic considerations are confirmed by the rigorous estimates below.
	
The density $\rho(\rv)$ can for $\omega\leq \omega_{\rm h}$ (see Eq.\ (\ref{omegah})) simply be taken to be equal to the TF density $\rho^{\rm TF}(\rv)$ \eqref{tfm} (note that $\rho^{\rm TF}$  depends on $\omega$). For $ \omega>  \omega_{\rm h} $, however, $\rho^{\rm TF}$ vanishes in a \lq hole' of radius $ R_{\rm h} =1-{\rm const.}(\omega)^{-1}$ (see Eq.\  (\ref{holeradius})) and
$\sqrt{ \rho^{\rm TF}}$ does not have  finite kinetic energy. Hence it is necessary in this case to regularize $\rho^{\rm TF}$ near the boundary of the hole. At the same time one has to take care that the TF energy of the regularized density remains close to $E^{\rm TF}$. Both conditions are met if we put
\beq
    	\label{rhoeps}
        	\rho(r) = 
			\lf\{
			\begin{array}{ll}
        	       		0						&	\mbox{if} \:\:  r\leq R_{\rm h}, \\
            			\tfma(\rtf+\Omega^{-1})\Omega^2(r-\rtf)^2	&   	\mbox{if} \:\: \rtf \leq r \leq \rtf+\Omega^{-1},  \\
            			\rho^{\rm TF}(r)   				&   	\mbox{otherwise}.
			\eay
		\ri.
\eeq
Thus the regularized density is equal to $\rho^{\rm TF}$ except in an annulus of thickness $\Omega^{-1}$ around the hole, where it increases quadratically with the distance from the hole. The latter property ensures finiteness of the kinetic energy.
We also note  that, by \eqref{holeradius} and \eqref{TFdensoutside}, $ \tfma(\rtf+\Omega^{-1})=O(\eps^2\Omega) $, so that, for any $ \rv \in \bo $,
\beq
	\label{rhoeps2}
	\rho(r)=\tfm(r) + O(\eps^2\Omega).
\eeq

We now collect some  simple  estimates that are needed in the proof of the upper bound. In the following  $C$ will stand for a positive, finite constant that may differ from line to line but is independent of $\Omega$ and $\eps$.

 First, note that $\rho^{\rm TF}\leq C (\omega+1)$ while the area of the support of $\rho^{\rm TF}$ is 
$C (\omega+1)^{-1}$.  The density of vortices is $ \Omega/2\pi$ and the area of each vortex disc is $\pi t^2$. Also, $\eps^2\Omega=o(1)$ by assumption. We thus have, using \eqref{vortinside} and \eqref{rhoeps2}, 
\beq 	
	\int\rho_{}\xi_{}^2=\int\rho_{}-\int\rho_{}(1-\xi_{}^2)\geq \int \rho^{\rm TF}-
\eps^2-C\,\Omega\cdot{t}^2\geq 1-O(t^2\Omega).\eeq
Hence the normalization constant satisfies
\beq \label{normconstant}c_{}\leq1+C\,t^2\Omega.\eeq
Likewise, using that $|\nabla\xi_{}|=t^{-1}$ in each vortex disc and zero outside the union of the discs, while the number of vortices in the support of $\rho_{}$ is $\leq (\omega+1)^{-1}\Omega$, 
\beq
	\label{xibound}
	\lf\|\sqrt{ \rho_{}}\,\nabla \xi_{} \ri\|^2_2 \leq C (\omega+1)\cdot t^{-2}\cdot (\omega+1)^{-1}\Omega\cdot t^2=C\,\Omega.
	\eeq
Next we consider, for $\omega>\omega_{\rm h}$, i.e., $R_{\rm h}>0$, 
	\beq
	\label{rhobound1}
	\Vert \xi_{}\nabla\sqrt{\rho_{}}\Vert_2^2\leq \frac{1}{4} \int\frac{|\nabla\rho_{}|^2}{\rho_{}}\leq
	\frac{1}{4} \int_{r< R_{\rm h}+\Omega^{-1}}\frac{|\nabla\rho_{}|^2}{\rho_{}}+
	\frac{1}{4} \int_{r\geq R_{\rm in}+\Omega^{-1}} 
		 \frac{|\nabla\rho^{\rm TF}|^2}{\rho^{\rm TF}}.\eeq
		By \eqref{rhoeps} and \eqref{rhoeps2} first term is bounded by $C\,\Omega\cdot (\eps^2\Omega )$. Using  \eqref{tfm}  we obtain
\beq
	\label{rhobound2} 
	\int_{r\geq R_{\rm h}+\Omega^{-1}} 
	\frac{|\nabla\rho^{\rm TF}|^2}{\rho^{\rm TF}}=  C\,(\eps\Omega)^2\int_{R_{\rm h}+\Omega^{-1}}^1\frac {r^3\,dr}{r^2-{R_{\rm h}}^2}\leq 
C\,(\eps\Omega)^2\int_{\Omega^{-1}}^{\omega^{-1}}\frac {du}{u}  \leq  C\,(\eps^2\Omega)\cdot\Omega|\log\eps|.  
\eeq
The above estimate shows that the closer $ \Omega $ is  to $ \eps^{-2} $, the larger is the kinetic contribution of the profile $ \sqrt{\tfm} $ and for $ \Omega \sim( \eps^2 |\log\eps|)^{-1} $ is becomes of the same order as the other remainders, i.e., $ \sim \Omega $.
		 For $\omega\leq \omega_{\rm h}$, on the other hand, $\rho=\rho^{\rm TF}$, and  $\nabla\sqrt{\rho}$ is uniformly bounded in $\omega$, so $\Vert \xi_{}\nabla\sqrt{\rho_{}}\Vert_2^2\leq C$ in this case.		 

Because $g_{}$ is a phase factor while $\sqrt{\rho_{}}$ and $\xi_{}$ are real-valued functions we have
\bml{
	\label{basic}
	\lf| \lf( \nabla-i\vec A_{} \ri) (\sqrt{\rho_{}} \: \xi_{}\;\phase ) \ri|^2 = \lf| \nabla \lf( \sqrt{\rho_{}} \: \xi_{} \ri) \ri|^2 + \xi^2\rho\lf| \lf( \nabla-i\vec A_{} \ri) \phase \ri |^2 \leq	\\
	2\xi_{}^2 \lf| \nabla\sqrt{\rho_{}} \ri|^2 + 2\rho_{} \lf| \nabla \xi_{} \ri|^2+  \xi_{}^2 \rho_{}\lf| \lf( \nabla - i\vec A_{} \ri ) \phase \ri|^2.}
We now obtain,  using Eqs.\ \eqref{normconstant}--\eqref{rhobound2}, 
\bml{
	\label{simplecomp}
	\gpf[\trial] -  \tff [ | \trial |^2 ] = c_{}^2 \int_{\bo} \diff \rv \: \lf| \lf( \nabla-i\vec A_{} \ri) \lf( \sqrt{\rho_{}} \: \xi_{}\;\phase \ri) \ri|^2 \leq	\\
	c_{}^2 \int_{\mathcal B_1} \diff \rv \: \xi_{}^2 \rho_{} \lf| \lf( \nabla-i\vec A_{} \ri) g_{} \ri|^2 + 2 c_{}^2 \int_{\mathcal B_1} \diff \rv \: \xi_{}^2 \lf |\nabla\sqrt{\rho_{}} \ri|^2 + 2c_{}^2 \int_{\mathcal B_1} \diff \rv\: \rho_{} \lf| \nabla\xi_{} \ri|^2 \leq	\\
	\lf( 1 + O( t^2\Omega) \ri) \int_{\bo} \diff \rv \: \xi_{}^2 \tfm \left| \left( \nabla - i \vec{A_{\eps}} \right) \phase \right|^2 + C\lf\{ \Omega + \eps^2 \Omega^2 |\log\eps| \ri\}.}	
	The estimate  for the vortex kinetic energy $ \int_{\mathcal B_1} \diff \rv \: \xi_{}^2\,\tfm |( \nabla - i \magnp ) \: \phase |^2$ is given in the following Proposition.

\begin{pro}[Vortex kinetic energy]
\mbox{}	\\
If  $\eps\to0$, and $1\ll \Omega\ll 1/\eps^2$, then
\beq 
	\label{phasebound}
        \int_{\mathcal B_1} \diff \rv \:\xi_{}^2\; \tfm \left| \left( \nabla - i \vec A \right) \: \phase \right|^2 \leq\half {\Omega}\,{|\log (t^2\Omega)|} + O(\Omega)+ O(\Omega\, (\eps^2\Omega)^{1/2}|\log (t^2\Omega)|).
\eeq
\end{pro}

\begin{proof}
	The idea behind the proof is the electrostatic analogy that was already mentioned in the Introduction and is made precise in Eq.\ (\ref{electricfield}) below and the considerations following it. The vortices, i.e., the singularities  of the phase factor $g$, play the role of unit charges while the vector potential corresponds, after a conformal transformation,  to the electric field of a uniform charge distribution. The density of the vortices is chosen in such a way that the field from the uniform charge distribution is compensated as far as possible and this requires in particular that each unit cell $Q_{}^i$ has total charge zero. If the unit cells were rotationally symmetric there would be no interaction between them by Newton's theorem. Complete rotational symmetry is, of course, not possible, but the closest approximation to it among the regular lattices is a lattice with hexagonal cells, i.e., a triangular arrangement of the vortices,  that gives the lowest electrostatic interaction energy. However, the difference between the three possible types of unit cells, triangular, rectangular and hexagonal does not show  up in  the  term of order $\Omega\,|\log(t^2\Omega)|$  but only in higher order corrections to this contribution.

To formalize these ideas we note first that $ | ( \nabla - i \magnp ) \: \phase |^2=|\nabla\phi_{}-\vec A_{}|^2$ where $\phi_{}=\sum_i\arg (\zeta-\zeta_i)$ is the
phase of $g_{}$. The conjugate harmonic function  
\beq
\tilde \phi_{}(\vec r)=\sum_i\log|\vec r-\vec r_i|
\eeq
 satisfies \beq\nabla\phi_{}=\nabla_r\tilde\phi_{}\,\vec e_\vartheta-\nabla_\vartheta\tilde\phi_{}\,\vec e_r\eeq
 where $\nabla_r=\vec e_r\cdot\nabla=\partial/\partial r$ and $\nabla_\vartheta=\vec e_\vartheta\cdot\nabla=r^{-1}\partial/\partial\vartheta$.
 With $\vec A_{}=A_{}(r)\vec e_\vartheta$ we thus have
 \beq 	
	\lf| \nabla \phi_{} - \magnp \ri|^2 = \lf| \nabla_\vartheta \tilde{\phi}_{} \ri|^2 + \lf| \nabla_r \tilde{\phi}_{} - A_{} \ri|^2 = \lf| \nabla_\vartheta \tilde{\phi}_{} \, \vec{e}_\vartheta +\nabla_r \tilde\phi_{} \, \vec e_r-A_{}\,\vec e_r \ri|^2= \lf| \nabla \tilde \phi_{}-A_{}\,\vec e_r \ri|^2.
\eeq
 We now define
 \beq\label{electricfield} \vec E(\vec r)=\nabla \tilde \phi_{}(\rv)-A_{}(r)\vec e_r\eeq
 and note that $\nabla \tilde \phi_{}=\sum_i(\rv-\rv_i)/|\rv-\rv_i|^2$ can be regarded as the electric field generated by point charges localized at the positions of the vortices, while $A_{}(r)\vec e_r
 =(\Omega/2)r \vec e_r$ is the field generated by a uniform charge density  of magnitude $\Omega/2\pi=|Q_{}^i|^{-1}$.
We can thus write $\vec E_{}(\rv) =\sum_i\vec E_{i}(\rv)=\sum_i\nabla\Phi_{ i}(\rv)$
with
\beq 
	\Phi_{i}(\rv) = \int_{\bo} \diff \rv^{\prime} \: \sigma_{i}(\rv')\log|\rv-\rv'| 
\eeq
and
\beq\label{chargedistr} \sigma_{i}(\rv')=\delta(\rv'-\rv_i)-|Q_{}^i|^{-1}\chi_{i}(\rv')\eeq
where $\chi_{i}$ is the characteristic function of the cell $Q_{}^i$.
\newline
By a transformation of variables, writing $\rv={\Omega}^{-1/2} \xv$, we map the cells $Q^i$ of side length $\ell\sim \Omega^{-1/2}$ onto cells $Q^i_1$ of side length $O(1)$. The characteristic function of $Q^i_1$ is denoted by $\chi_{i,1}(\xvp)$ and we use the index 1 also for the charge densities, electric fields and potentials generated by the cells $Q^i_1$. 
We can  then write
\beq \sigma_{i}(\rv)=\Omega\left[ \delta( \xvp -\xv_i)-|Q_1^i|^{-1}\chi_{i,1}(\xvp)\right]=\Omega\,\sigma_{i,1}(\xvp)\eeq
and
\beq\label{scaledE} E_{i}(\rv)=\Omega^{1/2}E_{i,1}(\xv)\eeq
where
\beq E_{i,1}(\vec x)=\nabla\int_{\bo} \diff \xvp \: \sigma_{i,1}(\xvp)\log|\xv-\xvp| =\nabla\Phi_{i,1}(\vec x).\eeq
Consider now the cell $Q_1^0$ centered at the origin.
The multipole expansion of $\Phi_{0,1}(\vec x)$ for $\vec x\notin Q_1^0$ is
\beq \Phi_{0,1}(\vec x)=q\log|\vec x|-\sum_{k=1}^\infty \frac{C_k\cos(k\vartheta)+S_k\sin(k\vartheta)}{|\vec x|^k}\eeq
with 
\bml{
	q=\int _{Q_1^0} \diff \xvp \: \sigma_{0,1}(\vec x'),	\\	
	C_k=k^{-1}\int_{Q_1^0} \diff \xvp \: \sigma_{0,1}(\vec x')|\vec x'|^k \cos(k\vartheta'), \qquad	S_k=k^{-1}\int_{Q_1^0} \diff \xvp \: \sigma_{0,1}(\vec x')|\vec x'|^k \sin(k\vartheta').}
By neutrality of the charge distribution \eqref{chargedistr} it is clear that $q=0$ and by symmetry of the unit cell it is also clear that there is no dipole moment, i.e., $C_1=S_1=0$. We conclude that $\Phi_{0,1}(\vec x)$ decays at least as $|\vec x|^{-2}$ and the corresponding  field $\vec E_{0,1}(\vec x)$  decays at least as $|\vec x|^{-3}$. For square or hexagonal cells it decreases even faster.

All cells $Q^i$ are obtained by translations and scaling from the cell $Q_1^0$. From the considerations above (note, in particular, Eq.\ \eqref{scaledE}) we can thus conclude that if two of the original cells,  $Q^i$ and  $Q^j$ have distance $O(\Omega^{-1/2}n)$ from each other, then the strength of the field $\vec E_j(\rv)$  for $\rv\in Q^i$ is at most $O(\Omega^{1/2}n^{-3})$. Since, for a fixed cell $Q^i$,  there are at most $O(n)$ cells at distance $O(\Omega^{-1/2}n)$ from it, we can estimate for $\vec r\in Q^i$
\beq \lf| \vec E_{}(\rv)-\vec E_{i}(\rv) \ri| \leq \sum_{j\neq i} |\vec E_{j}(\rv)|
\leq{\rm const.}\,\Omega^{1/2}\sum_n n\cdot n^{-3}=O(\Omega^{1/2}).
\eeq
Writing \beq |\vec E_{}|^2 = |\vec E_{i}|^2+ 2(\vec E_{}-\vec E_{i})\cdot \vec E_{i}  +
|\vec E_{}- \vec E_{i}|^2\eeq
and using the simple bound  $\vec E_{i}(\rv)\leq |\rv-\rv_i|^{-1}$, we conclude that for $\vec r\in Q^i$
\beq 
	|\vec E_{}(\rv)|^2\leq | \vec E_{i}(\rv) |^2+ {\rm const.} (\Omega^{1/2}|\rv-\rv_i|^{-1}+\Omega)
\eeq 
and hence
\beq 
	\int_{Q_{}^i \setminus{\mathcal B}_{t}^i} \diff \rv \: |\vec E_{}(\rv)|^2 -
\int_{Q_{}^i \setminus{\mathcal B}_{t}^i} \diff \rv \: |\vec E_{i}(\rv)|^2 \leq {\rm const.} \int_{{t}}^{C\,\Omega^{-1/2}} \diff r \, r \: \lf( \Omega^{1/2}r^{-1}+\Omega \ri) =O(1)
\eeq
while
\beq
	\int_{{\mathcal B}_{t}^i} \diff \rv \: \xi_{}(\rv)^2|\vec E_{}(\rv)|^2-
\int_{{\mathcal B}_{t}^i} \diff \rv \: \xi_{}(\rv)^2|\vec E_{i}(\rv)|^2  \leq {\rm const.}\int_0^{t} \diff r \, r \, ( r/t)^2 (\Omega^{1/2}r^{-1} +\Omega) =O((t^2\Omega)^{1/2}).
\eeq
On the other hand, since $\vec E_{i}(\rv)\leq |\rv-\rv_i|^{-1}$,
  \beq
\int_{Q_{}^i \setminus{\mathcal B}_{t}^i}\diff \rv \: |\vec E_{i}(\rv)|^2\leq 2\pi \int_{{t}}^{C\,\Omega^{1/2}} \diff r \, r \: r^{-2} =
\pi|\log(t^2\Omega)|+O(1)
\eeq
and
 \beq
\int_{{\mathcal B}_{t}^i} \diff \rv \: \xi_{}(\rv)^2|\vec E_{i}(\rv)|^2 \leq 2\pi\int_0^{{t}} \diff r \, r \, ( r(\Omega/\eps)^{1/2})^2  r^{-2} =
O(1).
\eeq
Putting all the estimates above together we obtain 
\beq
	\label{137}		
	\int_{\bo} \diff \rv \: \rho^{\rm TF}(\rv) \, \xi_{}(\rv)^2 |\vec E(\rv)|^2 \leq \lf(1+O((t^2\Omega)^{1/2}) \ri) \sum_i \sup_{\rv\in Q_{}^i } \rho^{\rm TF}(\rv) \lf( \pi|\log(t^2\Omega)|+O(1) \ri).
\eeq

It remains to estimate the  Riemann approximation error
\beq
	\label{riemannerror}		
	\mathcal R\equiv|Q_{}^0| \sum_i \sup_{\rv\in Q_{}^i } \rho^{\rm TF}(\rv) -\int_{\bo} \diff \rv \,\rho^{\rm TF}(\rv)\leq	|Q_{}^0| \sum_i \bigg\{ \sup_{\rv\in Q_{}^i } \rho^{\rm TF}(\rv)-\inf  _{\rv\in Q_{}^i } \rho^{\rm TF}(\rv) \bigg\}.
\eeq
We use here that $\Vert d\rho^{\rm TF}/dr\Vert_\infty\leq C (\eps\Omega)^2$ and that the number of cells $Q_{}^i$ that intersect the support of $\rho^{\rm TF}$ is bounded by $C\,\eps^{-1}(1+\Omega^{-1/2})$. Hence
\beq\mathcal R\leq C\Omega^{-1}\cdot\Omega^{-1/2}(\eps\Omega)^2\cdot \eps^{-1}(1+\Omega^{-1/2})=C\,(\eps^2\Omega)^{1/2}(1+\Omega^{-1/2}).
\eeq
It now follows that the right hand side of \eqref{137} is bounded by
\bml{
	(1+O((t^2\Omega)^{1/2}))\,(1+\mathcal R)\,|Q_{}^0|^{-1}
\lf( \pi|\log(t^2\Omega)|+O(1) \ri) =	\\
	\half\Omega\,|\log(t^2\Omega)| + O(\Omega)+O(\Omega\, (\eps^2\Omega)^{1/2}|\log (t^2\Omega)|).}
\end{proof}

To complete the proof of the upper bound we still need to estimate the difference between $\mathcal E^{\rm TF}[|\Psi_{}|^2]$ and $E^{\rm TF}=\mathcal E^{\rm TF}[\rho^{\rm TF}]$ and choose the radius $t$ of the vortex discs. 

The TF functional is
\beq 
	\mathcal E^{\rm TF}[|\Psi_{}|^2]=\eps^{-2}\int_{\bo} \diff\rv \: \left\{ |\Psi_{}|^4 - \frac{(\eps\Omega)^2 r^2 |\Psi_{}|^2}{4} \right\}.
\eeq
We consider the two terms separately. For the nonlinear interaction term we use that $c=1+O(t^2\Omega)$ and $\xi^2\rho\leq\rho^{\rm TF}$ to obtain
\beq
	\eps^{-2}\int_{\bo} \diff \rv \:  |\Psi|^4 = \eps^{-2} \int_{\bo} \diff \rv \: (c^2 \xi^2\rho)^2 \leq \frac{1+C\,t^2\Omega}{\eps^2} \int_{\bo} \diff \rv \:
\left(\rho^{\rm TF}\right)^2 =	\eps^{-2} \int_{\bo} \diff \rv \: \left(\rho^{\rm TF}\right)^2+\text{\rm remainder}.
\eeq
 Since $\rho^{\rm TF}\leq C\,(\eps\Omega+1)$ and $\int\rho^{\rm TF}=1$,  the remainder is
 \beq
 \frac{t^2\Omega}{\eps^2}\int_{\bo} \diff \rv \:
\left(\rho^{\rm TF}\right)^2\leq C \frac{t^2\Omega}{\eps^2}(\eps\Omega+1)
 \eeq
 and this has to be small compared to $\Omega|\log(t^2\Omega)|$.
 If $\Omega\lesssim 1/\eps$ this is clearly satisfied for $t=\eps$.  On the other hand, if $\eps\Omega\gg 1$ we can take $t=(\eps/\Omega)^{1/2}$ Note that for $\Omega\sim 1/\eps$ both choices coincide.
 
 The centrifugal contribution can by partial integration and using the normalization of $\Psi$ be written
\beq 
	-\frac{\Omega^2}4\int_{\bo} \diff \rv \: r^2|\Psi|^2=-\frac{\pi \Omega^2}{2}+\pi{\Omega^2}\int_0^1 \diff r \: r \: \Phi(r),
\eeq
 with 
 \beq\Phi(r)= \int_0^r \diff r^{\prime} \: r^{\prime} \: |\Psi(r^{\prime})|^2.\eeq
 Likewise,
  \beq -\frac{\Omega^2}4\int_{\bo} \diff \rv \: r^2\rho^{\rm TF}=-\frac{\pi \Omega^2}{2}+\pi{\Omega^2}\int_0^1 \diff r \: r \: \Phi^{\rm TF}(r),\eeq
   with 
 \bdm
	\Phi^{\rm TF}(r)=\int_0^r \diff r^{\prime} \: r^{\prime} \: \rho^{\rm TF}(r^{\prime}).
\edm 
From \eqref{rhoeps} and \eqref{normconstant} we obtain
 \beq\Phi(r)\leq\Phi^{\rm TF}(r)+C\,t^2\Omega.\eeq
 Moreover, the support of $\Phi$ as well as $\Phi^{\rm TF}$ has area $\leq (\eps\Omega+1)^{-1}$. Hence 
  \beq
  {\Omega^2}\int_0^1 \diff r \: \left\{\Phi^{\rm TF}(r)-\Phi(r)\right\} \leq C\,\Omega^2\cdot t^2\Omega
  (\eps\Omega+1)^{-1}.
  \eeq
  If $\eps\Omega$ is bounded and $t=\eps$, this is bounded by $C\,\Omega\cdot
  (\eps\Omega)^2\leq C\,\Omega$. If $\eps\Omega\gg 1$, we take $t^2=\eps/\Omega$ and obtain again $C\,\Omega$ as bound. 
\newline
We summarize the findings in the following

\begin{pro}[Energy upper bound]
\mbox{}	\\
For $\eps\to 0$ and $1\ll\Omega\lesssim 1/\eps$ we have
\beq\label{upperbound1} E^{\rm GP}\leq E^{\rm TF}+\half \Omega |\log(\eps^2\Omega)| + O(\Omega),
\eeq
and for $1/\eps\lesssim \Omega\ll 1/\eps^2$
\beq\label{upperbound2}
E^{\rm GP}\leq E^{\rm TF}+\half \Omega |\log\eps| + O(\Omega) + O(\Omega\,(\eps^2\Omega)^{1/2}|\log \eps|).
\eeq
\end{pro}

As we will see in the next section, the upper bounds are matched by corresponding lower bounds only in the parameter range $|\log\eps|\ll \Omega\ll 1/(\eps^2|\log\eps|)$. In fact, for 
$\Omega\lesssim |\log \eps|$ there are only finitely many vortices \cite{AftaDu, Igna1,Igna2, RD} and the upper bound
\eqref{upperbound1} is too large. For $ 1/(\eps^2|\log\eps|)\lesssim \Omega\ll 1/\eps^2$, on the other hand, a trial function different from \eqref{trial function} gives lower energy than \eqref{upperbound2}. This is the trial function considered in \cite{CDY1} Eq.\ (3.36) for $\Omega\gg \eps^{-1}$ that corresponds to a {\it \lq giant vortex'} where all the vorticity is concentrated at the center and the support of $\rho^{\rm TF}$ is vortex free. In fact, for such a trial function the next correction to the TF energy is $O(1/\eps^2)$, cf.\ \eqref{energyasyu}, and this is smaller than $\Omega |\log\eps|$ in the parameter range. This transition at $\Omega\sim 1/(\eps^2|\log\eps|)$ can also be understood by the following heuristic argument, employing the electrostatic analogy: For $\Omega\gg 1/\eps$ the number of cells in the support of $\rho^{\rm TF}$ is $\sim 1/\eps$.  Without vortices each cell has unit \lq charge', originating from the vector potential,  and the mutual interaction energy of the cells is of the order $1/\eps^2$. Putting a vortex in each cell neutralizes the charge so that the interaction energy becomes negligible, but instead there is an energy cost of order $\Omega|\log\eps|$ due to the vortices. Equating these two energies leads to $\Omega\sim 1/(\eps^2|\log\eps|)$ as the limiting rotational velocity above which the ansatz \eqref{trial function} is definitely not optimal.

It should be noted that also for $1/\eps\lesssim \Omega\ll1/(\eps^2|\log\eps|)$ one could for the upper bound replace the distribution of the vorticity on a lattice within the \lq hole'  by a single phase factor corresponding to a giant vortex at the origin, but in order to obtain the correction beyond the TF term the support of $\rho^{\rm TF}$ can not be vortex free. The detailed vortex distribution of the true minimizer of the GP energy functional is, however, an open question.

\section{Energy Lower Bound}

The lower bound to the GP ground state energy $ \gpe $  will be proved by a step-by-step reduction to the lower bound of the energy of a Ginzburg-Landau (GL) energy function for which results of \cite{Sa, SS1, SS2} can be employed.
As a preparation we first prove a bound on the GP minimizers in terms of the TF density:

\begin{lem}[Upper Bound for $ \lf|\gpm\ri| $]
	\label{boundedness}
	\mbox{}	\\
	For $ \eps \to 0 $ and $ |\log\eps| \ll \Omega \ll (\eps^2|\log\eps|)^{-1} $,
	\beq
	\label{bounded}
		\lf\| \gpm \ri\|^2_{\infty} \leq \tfm(1) (1 + o(1)).
	\eeq
\end{lem}

\begin{proof}
	Setting $ U_{} \equiv \lf| \gpm \ri|^2 $, we first note that the upper bound    		\eqref{upperbound2} and the trivial lower bound $E^{\rm GP}\geq E^{\rm TF}$ imply the convergence of $ U_{} $ to $ \tfm $ in $L^2$-norm. Indeed, using the simple bound $ 2 \tfm \geq \eps^2\tfchem + {\omega^2 r^2}/{4} $, the $L^1$ normalization of $ U_{} $, and the identity $ \tfchem = \tfe + \eps^{-2}\| \tfm \|^2_2 $, we have
	\begin{eqnarray}
	\label{l2convergence}
		\int_{\bo} \diff \rv \: \lf( U_{} - \tfm \ri)^2 \leq & \displaystyle{\int_{\bo}} \diff \rv \: \lf[ U_{}^2 - \tfchem U_{} - \frac{\omega^2 r^2 U_{}}{4} + {\tfm}^2 \ri] =	
		\eps^{2}\left( \tff\lf[ U_{} \ri] - \tfe\right)  \\ 
		&\leq  \eps^2\left( \gpe - \tfe\right)\leq C\,\Omega\eps^2|\log\eps|=o(1),
	\end{eqnarray}
	by \eqref{upperbound1} and \eqref{upperbound2} and the conditions on $ \Omega $.
 As a consequence
	\beq
		\lf\| U_{} \ri\|_2^2 - \lf\| \tfm \ri\|_2^2 =2 \int_{\bo} \diff \rv \: \tfm \lf( U_{} - \tfm \ri)+\int\diff \rv \:\lf(U-\tfm\ri)^2 \leq  \tfm(1)^{1/2} \,o(1)	
	\eeq
	where we have used the Schwarz inequality and the  trivial bound $ \| \tfm \|_2^2 \leq \Vert\tfm\Vert_\infty=\tfm(1)$, which follows from the $L^1$ normalization of $ \tfm $. Now $\tfm(1) \geq C (\omega+1)$ and theqrefore
	\beq\lf\| U_{} \ri\|_2^2 - \lf\| \tfm \ri\|_2^2\leq o(1)\,\tfm(1).
	\eeq
	Since $ \eps^2 (\chem - \tfchem) = \eps^2 (\gpe - \tfe) + \lf\| U_{} \ri\|_2^2 - \lf\| \tfm \ri\|_2^2 $ we thus have
	\beq
		\eps^2 (\chem - \tfchem)  \leq o(1)\, \tfm(1).
	\eeq
	Now acting as in the proof of Proposition 2.4 in \cite{CDY1}, we obtain from the variational equation \eqref{GP Equation} 	
	\bml{
		\label{subh}	
		- \frac{1}{2} \Delta U_{} \leq \lf[ \eps^2 \chem + \frac{\omega^2}{4} - 2 U_{} \ri] \frac{U_{}}{\eps^2} \leq \lf[ \eps^2 \lf( \chem - \tfchem \ri) + 2\lf( \tfm(1) -  U_{} \ri) \ri] \frac{U_{}}{\eps^2} \leq	\\
		2 \lf[ (1 + o(1)) \tfm(1) - U_{} \ri]  \frac{U_{}}{\eps^2},}
	by \eqref{tfchem} and the above estimate for $ \eps^2 (\chem - \tfchem ) $. At the maximum of $U$ the left hand side of \eqref{subh} is nonnegative and thus \eqref{bounded} holds.\end{proof}

We now proceed with the proof of the lower bound. The first step  is the extraction of the TF profile $ \tfm $ from the GP minimizer $ \gpm $, i.e., the ansatz $ \gpm = \sqrt{\tfm} u_{} $, which, one the one hand, allows to get rid of the leading order term in the energy asymptotics and, on the other hand, implies that $ u_{} $ minimizes a weighted GL functional. Unfortunately such a factorization is well defined only if the TF profile $ \tfm $ does not vanish inside $ \bo $ , i.e., for $ \omega < \omega_{\rm h} $. In order to get rid of this problem we first restrict the integration domain in the GP functional and set
\beq
	\label{tf support h}
	\supph \equiv \lf\{ \rv \in \bo \: \big| \: \tfm(r) \geq \omega |\log\delta_{}|^{-1} \ri\},
\eeq
with
\beq
	\label{delta}
	\delta_{} \equiv \eps^2 \Omega |\log\eps| \ll 1,
\eeq 
by \eqref{cond omega}. Note that
\beq
	\label{prop delta}
	|\log\delta| \leq C |\log\eps|,
\eeq
since $ \delta \gg \eps^2 |\log\eps|^2 $, by \eqref{cond omega}, so that $ 0 \geq \log\delta \geq \log(\eps^2|\log\eps|^2) \geq C \log\eps $. 
Note that $ \rin^2 + \omega^{-1} |\log\delta_{}|^{-1} < 1 $, since $ \rin^2 = 1 - C \omega^{-1} $ and $ |\log\delta| \gg 1 $, so, by \eqref{TFdensoutside}, the set $\supph$ is not empty.
Moreover, if $ \omega/ \omega_{\rm h}  $ is sufficiently small then the set $ \supph $ coincides with the whole trap $ \bo $ since, in that case, $ \tfm(r) \geq C > 0 $ for any $ \rv\ \in \bo $.
\newline
For $ \Omega $ and $\eps$ satisfying \eqref{cond omega} we now define for $ \rv \in \supph $
\beq
	\label{ext tf profile}
	\, u_{}(\rv)\equiv \gpm(\rv)\, {\tfm(r)}^{-1/2} .
\eeq
This is a smooth function with $ \lf| u_{} \ri|^2 \leq C |\log\delta_{}|  $ because of Lemma \ref{boundedness}.  Adding the kinetic energy term to both sides of \eqref{l2convergence} we obtain, exploiting the nonnegativity of the integrand,
\beq	
	\gpe \geq {\tfe} + \int_{\supp} \diff \rv \: \lf\{ \lf| \lf( \nabla - i \magnp \ri) \gpm  \ri|^2 + \eps^{-2} \lf( \tfm - \lf| \gpm \ri|^2 \ri)^2 \ri\}.
\eeq
Introducing the weighted GL-type functional
\beq
	\label{weighted GL h}
	\tgpf \lf[ u_{} \ri] \equiv \int_{\supph} \diff \rv \: \tfm(r) \lf\{ \lf| \lf( \nabla - i \magnp \ri) u_{} \ri|^2 + \eps^{-2} \tfm(r) \lf( 1 - \lf| u_{} \ri|^2 \ri)^2 \ri\},
\eeq  
we thus obtain, since $ \supph \subset \supp $,
\bml{
	\gpf \lf[ \gpm \ri] - {\tfe} - \tgpf \lf[ u_{} \ri] \geq  \frac{1}{2} \int_{\supph} \diff \rv \: \nabla \tfm \cdot \nabla \lf| u_{} \ri|^2 \geq \const \omega^2 \int_{\supph} \diff \rv \: \rv \cdot \nabla \lf| u_{} \ri|^2 \geq \\ 
	- \const \omega^2 \int_{\supph} \diff \rv \: \lf| u_{} \ri|^2 \geq - \const \omega |\log\delta_{}|,}
which yields, by \eqref{prop delta},
\beq
	\label{ext leading h}
	\gpe \geq {\tfe} + \tgpf \lf[ u_{} \ri] - \const \omega |\log\eps|.
\eeq

According to \eqref{ext leading h} the correction to the leading term  $ \tfe$ can thus be estimated from below by a weighted GL energy $ \tgpf[u_{}] $, where the Lebesgue measure is replaced by $ \tfm(\rv)\, \diff \rv $. Compared with the usual GL setting there are two differences, however: The internal magnetic field $ \magnp $ is in our case fixed from the outset and the coupling parameter is $ \tfm(\rv)\, \eps^{-2} $, i.e., it depends on the TF density at each position. 

To deal with the latter point we decompose the integration domain into small cells: Let $\hat{ \latt} $ be the square regular lattice  
\beq
	\label{latticelb}
	\hat{\latt} \equiv \lf\{ \rv_i = (m \hat{\spac}, n \hat{\spac}), m,n \in \Z \: \big| \: \celli \subset \supph \ri\},
\eeq
where $ \celli $ denotes the lattice cell centered at $ \rv_i \in \hat{\latt} $ and
\beq
	\label{spacinglb}
	\sqrt{\frac{|\log\eps|}{\Omega}} \ll \hat{\spac} \ll \min\lf[1, \: \frac{1}{\omega|\log\delta_{}|} \ri].
\eeq
Note that the above conditions are compatible: Multiplying both side by $ \omega $ and assuming that $ \omega \geq |\log\delta_{}|^{-1} $, \eqref{spacinglb} becomes $ \sqrt{\delta_{}} \ll \omega \hat{\spac} \ll |\log\delta_{}|^{-1} $, which can always be fulfilled since $ \delta_{} \ll  1 $ by definition. Note also that the lattice spacing is much larger than the one chosen in the upper bound proof, where the lattice constant was $\ell\sim\Omega^{-1/2}$. By the lower bound  on $\hat\ell$ each lattice cell can be expected to contain a large number of vortices which turns out to be helpful for estimating the energy. The upper bound on $\hat\ell$ guarantees that $ \hat{\spac} $ is much smaller than the width of $ \supph $, which is of order $ (\omega+1)^{-1} $. This is useful for the extraction of  the TF profile.

By \eqref{tfm}, $ \tfm(r) \geq \tfm(r_i) (1 - {O}(\hat{\spac} \omega |\log\delta_{}|)) $, for any $ \rv \in \celli $, so that the above inequalities imply
\bml{
	\label{cell decomposition}
	\tgpe \lf[ u_{} \ri] \geq \sum_{\rv_i \in \latt} \int_{\celli} \diff \rv \: \tfm(r) \lf\{ \lf| \lf( \nabla - i \magnp \ri) u_{} \ri|^2 + \eps^{-2} \tfm(r) \lf( 1 - \lf|u_{}\ri|^2 \ri)^2 \ri\} \geq \\
	(1 - o(1)) \sum_{\rv_i \in \latt} \tfm(r_i) \: \gpfi [u],}
with 
\beq
	\label{GL functional}
\gpfi [ u] \equiv \int_{\celli} 
	\diff \rv \: \lf\{ \lf| \lf( \nabla - i \magnp \ri) u_{} \ri|^2+  \eps^{-2} \tfm(r_i) \lf( 1 - |u|^2 \ri)^2 \ri\}.
\eeq
Now the analogy with the GL functional is made explicit, since, except for the coupling parameter which still contains $ \tfm $, the functional $ \gpfi $ is precisely the GL energy functional
\beq
	\label{GL comp}
	\glf \big[ u, \vec{A}'\big] = \int_{\celli} \diff \rv \:  \lf\{ \lf| \lf( \nabla - i \vec{A}' \ri) u \ri|^2 + \lf| \nabla \wedge \vec{A}' - \vec h_{\mathrm{ex}} \ri|^2 + \eps^{-2} \tfm(r_i) \lf( 1 - \lf| u \ri|^2 \ri)^2 \ri\}
\eeq
 evaluated at $ (u_{}, \magnp) $ with an external magnetic field $ \vec h_{\mathrm{ex}} = \Omega \vec e_z$.
It is clear that the GL energy \beq
\label{GL energy} 
E^{\rm GL}\equiv \inf_{u, \vec{A}'} \glf [u, \vec{A}']
\eeq
is a lower bound to the ground state energy of $\gpfi$
because  the configuration with the uniform internal magnetic field corresponding to $\vec A'=\vec A=\Omega \vec e_z\wedge\rv/2$ is only one among all possible configurations considered in the minimization of the GL functional.

We can now state the main estimate needed for the proof of the lower bound.

\begin{pro}[Lower bound inside cells]
	\label{lower bound gc}
	\mbox{}	\\
	For any $ \Omega $ satisfying \eqref{cond omega} and $ \eps $ sufficiently small, it is possible to find $ \hat{\spac} $ in such a way that \eqref{spacinglb} is fulfilled and  
	\beq
		\label{lb good cells}
		\gpfi\lf[ u \ri] \geq \frac{\Omega \hat{\spac}^2 |\log\gamma_{}|}{2}\lf(1 - o(1) \ri),
	\eeq
	where $ \gamma_{} \equiv \min [\eps, \eps^2\Omega] $.
\end{pro}

\begin{proof}
	The key point in the proof of \eqref{lb good cells} is a rescaling of $ \celli $ (together with the choice \eqref{spacinglb} of the lattice spacing), which allows to reduce the problem to the minimization of a GL functional in a different regime.
	We thus set $ \xv \equiv \hat{\spac}^{-1} (\rv - \rv_i) $, 
	\beq
		\label{rescaling}
		\tilde{u}_{}(\xv) \equiv u_{} \lf(\rv_i + \hat{\spac} \xv \ri), \qquad	\rmagnp(\xv) \equiv \hat{\spac} \magnp \lf( \rv_i + \hat{\spac} \xv \ri),
	\eeq
	where $ \rv_i $ stands for the center of $ \celli $. By such a change of coordinates in \eqref{GL functional}, we obtain
	\beq
		\label{rescaled GLF}
		\gpfi[u]=\gpfit\lf[ \tilde{u}_{} \ri] =  \int_{\ucell} \diff \xv \: \lf\{ \lf| \lf( \nabla - i \rmagnp \ri) \tilde{u}_{} \ri|^2 + \eps^{-2} \hat{\spac}^2 \tfm(r_i) \lf( 1 - \lf| \tilde{u}_{} \ri|^2 \ri)^2 \ri\},
	\eeq
	where $ \ucell $ is a unitary square centered at the origin. Note that the  rescaled vector potential $ \rmagnp $ is explicitly given by
	\beq
		\label{scaled vect}
		\rmagnp(\xv) = \frac{\Omega \hat{\spac} \vec{e}_z \wedge \rv_i}{2} +  \frac{\Omega \hat{\spac}^2 \vec{e}_z \wedge \xv}{2}, 
	\eeq
	and the corresponding magnetic field is 
	\beq
		\tilde{h}_{} \equiv \mathrm{curl} \rmagnp = \Omega \hat{\spac}^2.
	\eeq
	In the following we investigate the minimization of the functional $ \gpfit $: We first notice that, by gauge invariance, one can get rid of the constant term $ \Omega \hat{\spac} \vec{e}_z \wedge \rv_i / 2$ in \eqref{scaled vect}:
	\beq
		\label{gauged away}
		\inf_{\tilde u \in H^1(\ucell)} \gpfit[\tilde u] \geq \inf_{\tilde u \in H^1(\ucell)} \int_{\ucell} \diff \xv \: \lf\{ \lf| \lf( \nabla - i \hat{\spac}^2 \magnp(\xv) \ri) \tilde u \ri|^2 + \eps^{-2} \hat{\spac}^2 \tfm(r_i) \lf( 1 - \lf| \tilde u \ri|^2 \ri)^2 \ri\}.
	\eeq
	We now introduce a new infinitesimal parameter $ \epsilon $ defined as
	\beq
		\label{new par}
		\epsilon \equiv \frac{\eps}{\hat{\spac} \sqrt{\tfm(r_i)}} \leq C \eps \sqrt{\frac{\Omega |\log\delta_{}|}{\omega|\log\eps|}} \leq C \sqrt{\eps} \ll 1,  
	\eeq
	by \eqref{delta}, \eqref{tf support h} and \eqref{spacinglb}. It follows that
	\beq
		\label{scaledfunc}
		\gpfi [ u ] \geq \inf_{\tilde u \in H^1(\ucell)} \int_{\ucell} \diff \xv \: \bigg\{ \bigg| \bigg( \nabla - \frac{i \tilde h_{\mathrm{ex}} \vec{e}_z \wedge \xv}{2} \bigg) \tilde u \bigg |^2 + \epsilon^{-2} \left ( 1 - \lf | \tilde u \right|^2 \ri)^2 \bigg\},
	\eeq 
	for a magnetic field $ \tilde h_{\mathrm{ex}}$ satisfying the conditions 
	\beq
		\label{magn cond}
		|\log\epsilon| \ll \tilde h_{\mathrm{ex}} = \Omega \hat{\spac}^2  \ll \frac{1}{\epsilon^{2}}.
	\eeq
	Indeed, by \eqref{tf support h} and \eqref{new par},
	\beq
		\Omega \hat{\spac}^2 = \frac{\Omega \eps^2}{\epsilon^2 \tfm(r_i)} \leq \frac{\eps |\log\delta_{}|}{\epsilon^2} \ll \frac{1}{\epsilon^2},	\qquad	\Omega \hat{\spac}^2 \gg |\log\eps| \geq |\log\epsilon|,
	\eeq
	because $ 0 \geq \log\epsilon = \log\eps - \log(\hat{\spac}\sqrt{\tfm(r_i)}) $ and 
	\beq
		\hat{\spac} \sqrt{\tfm(r_i)} \ll \min \lf[ \frac{1}{\sqrt{|\log\delta_{}|}}, \: \frac{1}{\sqrt{\omega} |\log\delta_{}|} \ri] \ll 1,
	\eeq 
	which implies $ \log \epsilon \geq \log\eps $ and $ |\log\epsilon| \leq |\log\eps| $.
	\newline
	The functional on the right hand side of \eqref{scaledfunc}  is precisely  the GL functional on $\ucell$ with external magnetic field $ \tilde h_{\mathrm{ex}}\vec e_z  $ and parameter $\epsilon$, i.e.,
	\beq
		\tilde{\mathcal{E}}^{\mathrm{GL}} \lf[ \tilde u, \vec{A}' \ri] \equiv \int_{\ucell} \diff \xv \: \lf\{ \lf| \lf( \nabla - i \vec{A}' \ri) \tilde u \ri|^2 + \lf| \mathrm{curl} \vec{A}' - \tilde h_{\mathrm{ex}}\vec e_z \ri|^2 + \epsilon^{-2} \lf( 1 - \lf| \tilde u \ri|^2 \ri)^2 \ri\},
	\eeq
	evaluated on the configuration 
	\beq
		\lf( \tilde u \: ,\vec{A}' \ri) = \lf( u \, , {\tilde h_{\mathrm{ex}}(\epsilon) \vec{e}_z \wedge \xv}/{2} \ri).
	\eeq
		and \eqref{magn cond} corresponds to the GL regime where the external magnetic field is  between the first and the second critical fields. We can thus apply the lower bound for the GL functional proven in \cite{SS1}, Theorem 1.1 (note that in the definition of the GL functional given in \cite{SS1} there is overall factor $ 1/2 $), to get
	\beq	
		\gpfi [ u] \geq (1-o(1)) h_{\mathrm{ex}} \log \frac{1}{\epsilon \sqrt{h_{\mathrm{ex}}}} = (1-o(1))\frac{\Omega \hat{\spac}^2}{2} \log  \frac{\tfm(r_i)}{\eps^2 \Omega}  \geq \lf( 1 - o(1) \ri) \frac{\Omega \hat{\spac}^2 |\log\gamma_{}|}{2}, 
	\eeq  
	since $ \tfm(r_i) \geq \omega |\log\delta_{}|^{-1} $ inside $ \supph $, if $ \Omega \gtrsim \eps^{-1} $, and $ \tfm(r_i) \geq \const $, if $ \Omega \ll \eps^{-1} $.
\end{proof}	

\noindent
The proof of the lower bound to the GP energy $\gpe$ is now almost complete. Collecting the lower bounds inside all cells proven in the proposition above, we have
\beq
	\tgpe \lf[ u_{} \ri] \geq \frac{\Omega \hat{\spac}^2 |\log\gamma_{}|}{2} \sum_{\rv_i \in \latt} \tfm(r_i) (1-o(1)).
\eeq
The replacement of the Riemann sum by the integral can be done exactly as in \eqref{cell decomposition}: By the symmetry of the lattice cell and the $ L^1-$normalization of $ \tfm $,
\beq
	\sum_{\rv_i \in \latt} \tfm(r_i) \geq \frac{1}{\hat{\spac}^2} \lf( \int_{\cup_i \celli} \diff \rv \: \tfm(r) - \const \max[\hat{\spac}, \omega\hat{\spac}] \ri) \geq \frac{1 - o(1)}{\hat{\spac}^2},
\eeq
and we finally obtain\beq
	\gpe \geq \tfe + \frac{ \Omega |\log\gamma_{}|}{2} (1 -o(1)).
\eeq
Since $\gamma=\min [\eps, \eps^2\Omega]$ this gives the lower bounds in  
\eqref{energyub1} and \eqref{energyub2}. Note that the condition $\Omega\ll 1/(\eps^2|\log\eps|)$ entered in \eqref{delta}.

\section{Vorticity of GP Minimizers}

In this Section we prove Theorem \ref{distribution}, which is a consequence of the energy asymptotics in \eqref{energyub1} together with a similar result in GL theory.  Indeed we shall prove that one can associate to any GP minimizer a GL configuration satisfying certain energy bounds, which, by exploiting a result proven in \cite{SS1}, yield the uniform distribution of vorticity. The proof closely follows the analysis performed in Section 5 in \cite{SS1} and relies on Proposition 5.1 in this reference as a key ingredient.

It is appropriate to point out that
	Theorem \ref{distribution} is  a statement about the uniform distribution of the local winding numbers of $\Psi$  in the support of $\tfm$
	 but not about the nature of the singularties.  The energy considerations behind this result are not sufficient to exclude the occurrence of singularities that are not pointlike, e.g., lines of zeros of $\Psi^{\rm GP}$. While we expect that $\Psi^{\rm GP}$ contains only isolated vortices  in the parameter range \eqref{cond omega},  a proof of this has not been accomplished. The same is true  for the  corresponding question in GL theory (see, e.g., \cite{SS1, SS2}).

\begin{proof1}{Theorem \ref{distribution}}
	Exploiting the energy bounds proved before (see \eqref{upperbound1}, \eqref{ext leading h}, \eqref{cell decomposition} and \eqref{lb good cells}), we obtain, with $\hat\ell$ and $\hat{\mathcal L}$ as in \eqref{latticelb}-\eqref{spacinglb}, 
	\beq
		- o(1) \, \Omega \hat{\spac}^2 |\log(\eps^2\Omega)| \sum_{\rv_i \in \hat{\latt}} \tfm(r_i) \leq  \sum_{\rv_i \in \hat{\latt}} \tfm(r_i) \lf[ \gpfi[u_{}] - \frac{\Omega \hat{\spac}^2 |\log(\eps^2\Omega)|}{2} \ri] \leq C \Omega,
	\eeq
	so that, since the sum is performed over the lattice $ {\hat{\latt}} \subset \mathcal{T}_{} $, i.e., where $ \tfm \geq \omega |\log\delta|^{-1} $,
	\beq
		\label{restriction}
		\sum_{\rv_i \in \hat{\latt}} \tfm(r_i) \lf| \gpfi[u_{}] - \frac{\Omega \hat{\spac}^2 |\log(\eps^2 \Omega)|}{2} \ri| \leq g(\eps) \: \Omega \hat{\spac}^2 |\log(\eps^2 \Omega)|  \sum_{\rv_i \in \hat{\latt}} \tfm(r_i),
	\eeq
	for some $ g(\eps) \to 0 $, as $ \eps \to 0 $.
	\newline
	Now we can distinguish, as in \cite{SS1}, between good and bad cells, where the above inequality yields an upper (resp. lower) bound. The key point is that, if the definition of such cells is done in the appropriate way, the upper bound can be used to prove a uniform distribution of vorticity (inside good cells) and, at the same time, there are only few bad cells, i.e., their number is only a remainder with respect to the total number of cells. The final result would then be a simple consequence of the fact that cells cover $ \supp $ in the limit $ \eps \to 0 $.
	\newline
	We say that a cell $ \celli $ is a \emph{good cell}, if 
	\beq
	\label{good cell}
		\gpfi[u_{}] - \frac{\Omega \hat{\spac}^2 |\log(\eps^2 \Omega)|}{2} \leq \sqrt{g(\eps)} \: \Omega \hat{\spac}^2 |\log(\eps^2 \Omega)| ,
	\eeq
	while inside bad cells the inequality is reversed.
	\newline
	We can thus apply to any good cell Proposition 5.1 in \cite{SS1}, which implies the existence of a finite family of disjoint discs $ \bii $, $ i = 1,\ldots,k_{} $, such that the sum of all the radii is bounded by $ \Omega^{-1/2} $ and $ | u_{}| > 1/2 $ on $ \partial \bii $. Points 1, 2 and 3 in Proposition \ref{distribution} then easily follows. In particular point 2 follows from a simple bound on the total number of cells, i.e., $ \numb \ll \Omega^{-1} $, which is a consequence of the conditions \eqref{spacinglb} on $ \hat{\spac} $.
	\newline
	Furthermore, setting $ d_{}^i $ equal to the winding number of $ u_{} $ on $ \partial \bii $, which is also the winding number of $ |\gpm|^{-1} \gpm $ because $ \tfm > 0 $ inside $ \mathcal{T} $, we have 
	\beq
	\label{sum wind}
		2 \pi \sum d _\eps^i \geq \Omega \hat{\spac}^2 (1 - o(1)), \qquad	2\pi \sum |d_\eps^i| \leq \Omega \hat{\spac}^2 (1 + o(1)).
	\eeq
	The second estimate above in particular implies that, by \eqref{spacinglb}, the measure on the left hand side of \eqref{measure} is uniformly bounded in $ \eps $, which guarantees its weak convergence. It remains only to show that it converges to the uniform measure on $ \supp $. To this purpose we first have to show that the number of bad cells included in any given open set $ \mathcal{S} \subset \supp $ (independent of $\eps$, i.e., such that $ |\mathcal{S}| \geq C > 0 $) is, for $\eps\to 0$, much smaller than the total number of cells  in $ \mathcal{S}$.  In fact, since the area of $\mathcal S$ is positive and the diameter of the cells tends to zero for $\eps\to 0$, it  suffices to shows that this is true for $ \mathcal{S} = \supp $. Denote by $ \mathcal{I} $ 
	 the sets of indices $ i \in \N $ such that $ \celli $ is a good (resp. bad) cell, 
	 then, by definition of bad cell and \eqref{restriction}, 
	\beq	 
		N^{\mathrm{B}}_{} \Omega \hat{\spac}^2 |\log(\eps^2\Omega)| \sqrt{g(\eps)} \leq \sum_{i \in \mathcal{J}} \lf[ \gpfi[u_{}] - \frac{\Omega \hat{\spac}^2 |\log(\eps^2\Omega)|}{2} \ri] \leq \const N_{} \Omega \hat{\spac}^2 |\log(\eps^2\Omega)| g(\eps),
	\eeq
	where $ N_{}^{\mathrm{B}} $ denotes the number of bad cells
	  and $ \numb $ the total number of cells. As a consequence
	\beq
		N_{}^{\mathrm{B}} \leq \const \sqrt{g(\eps)} \numb,
	\eeq
	and, since $ g(\eps) = o(1) $, the number of bad cells is always much smaller than the total number of cells. The result can be easily extended to any set $ \mathcal{S} \subset \bo $ by observing that the upper and lower bounds to the energy applies to any open subset of $ \bo $.
	\newline
	Theqrefore, for any given open subset $ \mathcal{S} \subset \supp $, good cells exhaust the whole of $ \mathcal{S} $ as $ \eps \to 0 $, i.e., $ N_{}^{\mathrm{G}} \hat{\spac}^2 \to |\mathcal{S}| $. Now, collecting all the disc families inside good cells and setting 
	\beq
		\mu_{} \equiv \frac{2\pi}{\Omega} \sum d_{i,\eps} \delta\lf(\rv - \rv_{i,\eps} \ri),
	\eeq
	where $  \rv_{i,\eps} $ stands for the center of $ \bii $, one has, by the first estimate in \eqref{sum wind},
	\beq
	  	\mu_{}(\mathcal{S}) \geq  N_{}^{\mathrm{G}} \hat{\spac}^2 \geq (1-o(1)) |\mathcal{S}|,
	\eeq
	and similarly, by the second estimate in \eqref{sum wind},
	\beq
		\mu_{}(\mathcal{S}) \leq (1 + o(1)) N_{}^{\mathrm{G}} \hat{\spac}^2 \leq (1+o(1)) |\mathcal{S}|,
	\eeq
	which implies \eqref{measure}, since $ \mathcal{S} $ is arbitrary. 
\end{proof1}

\section{Conclusions} Within the framework of two-dimensional GP theory we
have evaluated exactly to subleading order the contributions of vorticity
to the energy of a rapidly rotating Bose-Einstein condensate in a finite,
flat trap. The results of the mathematical analysis lend support to the
physical picture of a large number of vortices that are arranged in a
triangular lattice at not too high rotational velocities but are
eventually replaced by a \lq giant vortex' with all the vorticity
located outside the bulk of the density at sufficiently fast rotation. It
would be desirable to substantiate this picture even further by
generalizing Theorem \ref{distribution} for $\Omega\gg \eps^{-1}$ and by
proving a lower bound to the energy matching the \lq giant vortex' upper
bound of \cite{CDY1} to subleading order for $\Omega\gtrsim
\eps^{-2}|\log\eps|^{-1}$. Further interesting open problems concern the
nature of the singularities of the GP minimizer, in particular the
exclusion of line singularities and the precise arrangement of the
vortices. This would in particular require energy estimates beyond the
subleading order considered here.

\vspace{1cm}

\noindent
\textbf{Acknowledgements:} MC gratefully acknowledges the hospitality of the Erwin Schr\"{o}dinger Institute (ESI). JY  thanks the Science Institute of the University of Iceland for hospitality. This work was supported by an Austrian Science Fund (FWF) grant P17176-N02.

\appendix
\section{The TF Energy and Density}

We collect here from  \cite{CDY1} some formulas  for the TF energy and density (in a slightly different notation).
\newline
Defining \beq\label{omegah}
	\omega_{\rm h}\equiv4/\sqrt{\pi},
\eeq 
we have
\beq
    	\eps^2\tfe = 
		\lf\{
		\begin{array}{ll}
			\displaystyle{\frac{1}{\pi}}-\frac{\omega^2}{8} - \frac{\pi \omega^4}{768},      	&   \mbox{if}	\:\:  \omega \leq \omega_{\rm h},    \\ 
			\mbox{}	&	\mbox{}	\\
            		- \displaystyle{\frac{\omega^2}{4}} \lf[ 1 - \frac{8}{3\sqrt{\pi} \omega} \ri ],   	&   \mbox{if} 	\:\: \omega > \omega_{\rm h},
		\eay
		\ri.
\eeq
\beq
	\label{tfm}
    	\tfm(r) = 
		\lf\{
		\begin{array}{ll}
			\displaystyle{\frac{1}{\pi}}+\frac{\omega^2}{16} - \frac{\omega^2}{8} (1-r^2),			&	\mbox{if} \:\: \omega \leq \omega_{\rm h},    \\ 
			\mbox{}	&	\mbox{}	\\
            		\left[ \displaystyle{\frac{\omega}{2 \sqrt{\pi}}}-\frac{\omega^2}{8} (1-r^2) \right]_+,    	&   	\mbox{if}	\:\: \omega > \omega_{\rm h},
		\eay
		\ri.
\eeq
where $[t]_+=t$, if $t\geq 0$, and 0 otherwise. The TF density $ \tfm $ can be as well expressed as 
\beq
	\tfm(r) = \frac{1}{2} \lf[ \eps^2 \tfchem + \frac{\omega^2 r^2}{4} \ri]_+,
\eeq
where the chemical potential $ \tfchem = \tfe + \eps^{-2} \| \tfm \|_2^2 $ is fixed by the normalization of $ \tfm $ and it is explicitly given by
\beq
	\label{tfchem}	
	\eps^2 \tfchem = 
			\lf\{
			\begin{array}{ll}
				\displaystyle{\frac{2}{\pi}} - \frac{\omega^2}{8},				&	\mbox{if} \:\: \omega \leq \omega_{\rm h},    \\ 
				\mbox{}	&	\mbox{}	\\
            			-\displaystyle{\frac{\omega^2}{4}} \lf[ 1 - \frac{4}{\sqrt{\pi} \omega} \ri],	&	\mbox{if} \:\: \omega > \omega_{\rm h}.
			\eay
			\ri.
\eeq

Note that, if $ \omega > \omega_{\rm h}$, a \lq hole' centered at the origin occurs in the TF minimizer, i.e.,
	$\tfm(r) = 0$ {\rm for all } $ r \leq \rin$  with
	\beq\label{holeradius}
		\rin \equiv \left({1 - \frac{\omega_{\rm h}}{ \omega}}\right)^{1/2}
	\eeq
the radius of the hole. For $\omega\geq \omega_{\rm h}$ and $R_{\rm h}\leq r\leq 1$ we can also write the density as
\beq\label{TFdensoutside}
\tfm(r)= \frac{\omega^2}8(r^2-R_{\rm h}^2).
\eeq

Note also that the behaviour of $ \tfe $ in the regimes $ \Omega \ll \eps^{-1} $ ($ \omega \to 0 $) and $ \Omega \gg \eps^{-1} $ ($ \omega \to \infty $) is respectively
\beq
	\label{TF asymptotics}
	\eps^2 \tfe = \frac{1}{\pi} - O(\omega^2), \qquad	\eps^2\tfe = - \frac{\omega^2}{4} (1 - O(\omega^{-1})).
\eeq

\end{document}